# Models for Spatially Resolved Conductivity of Rectangular Interconnects with Integrated Effect of Surface And Grain Boundary Scattering

Xinkang Chen, and Sumeet Kumar Gupta, Senior *Member,* IEEE

*Abstract*—Surface scattering and grain boundary scattering are two prominent mechanisms dictating the conductivity of interconnects and are traditionally modeled using the Fuchs-Sondheimer (FS) and Mayadas-Shatzkes (MS) theories, respectively.  In addition to these approaches, modern interconnect structures need to capture the space-dependence of conductivity, for which a spatially resolved FS (SRFS) model was previously proposed to account for surface scattering based on Boltzmann transport equations (BTE).

In this paper, we build upon the SRFS model to integrate grain-boundary scattering leading to a physics-based SRFS-MS model for the conductivity of rectangular interconnects. The effect of surface and grain scattering in our model is *not* merely added (as in several previous works) but is appropriately integrated following the original MS theory. Hence, the SRFS-MS model accounts for the interplay between surface scattering and grain boundary scattering in dictating the spatial dependence of conductivity. We also incorporate temperature (*T*) dependence into the SRFS-MS model. Further, we propose a circuit compatible conductivity model (SRFS-MS-C3), which captures the space-dependence and integration of surface and grain boundary scattering utilizing an analytical function and a few (three or four) invocations of the physical SRFS-MS model. We validate the SRFS-MS-C3 model across a wide range of physical parameters, demonstrating excellent agreement with the physical SRFS-MS model, with an error margin of less than 0.7%. The proposed SRFS-MS and SRFS-MS-C3 models explicitly relate the spatially resolved conductivity to physical parameters such as electron mean free path ($\lambda_0$), specularity of surface scattering (*p*), grain boundary reflectance coefficient (R), interconnect cross-section geometry and temperature (*T*).

## I. INTRODUCTION

Interconnects pose a significant challenge in sustaining the scaling trajectory of modern electronics, necessitating focused attention to overcome associated limitations [1]. The main challenge stems from an increase of the interconnect resistivity with dimension scaling due to worsening surface/grain boundary scattering.

This work was supported, in part, SRC/NIST-funded NEW materials for LogIc, Memory and InTerconnectS (NEWLIMITS) Center, JUMP2 CoCoSys center and NSF FuSe. (Corresponding author: Xinkang Chen.)

Xinkang Chen and Sumeet Kumar Gupta are with the School of Electrical and Computer Engineering, Purdue University, West Lafayette, IN 47907 USA (e-mail: chen3030@purdue.edu; guptask@purdue.edu).

Various innovative designs and alternative materials have been proposed as potential solutions to improve interconnect performance [2], [3]. However, interconnect scaling continues to be a critical issue, as various studies forecast that interconnects will become primary performance bottlenecks in advanced technology nodes due to scaling complexities [1], [4]. Consequently, a thorough understanding and accurate modeling of interconnect conductivity and scattering mechanisms are essential for identifying viable future interconnect candidates.

Surface scattering is a key mechanism that significantly contributes to the increased resistivity observed in interconnects as their cross-sectional area is reduced. The well-established Fuchs-Sondheimer (FS) theory [5] models surface scattering through fundamental physical equations. To cater to the needs of modern interconnect (specifically via) structures, we proposed a spatially resolved FS (SRFS) model derived from the Boltzmann transport equation (BTE) [6]. We had also utilized the SRFS model to propose a circuit-compatible conductivity model (the SRFS-C3 model) [7], which offers spatial resolution of conductivity and explicit relation to the interconnect material and geometry parameters. This approach provided a significant enhancement over current empirical models that predict space dependence of conductivity but lack physical insights.

In addition to surface scattering, grain boundary scattering is another critical scattering mechanism affecting interconnect resistivity. This phenomenon occurs when the movement of electrons is hindered by interfaces between individual crystalline grains within a material. These grain boundaries present a barrier leading to scattering of the charge carriers, thereby increasing the resistance of the material. One of the most widely used theory to model grain boundary scattering is the Mayadas-Shatzkes (MS) theory [8].

By combining the MS [8] and FS (Fuchs-Sondheimer) theories [5], the resistivity of a conductor with given physical parameters (such as dimensions and electron mean free path ($\lambda_0$)) can be determined. Several studies [9], [10] have utilized these theories to estimate resistivity in scaled technologies. Many previous works combine the FS and MS conductivities using Matthiessen's rule i.e. by simply adding the individual components. While this assumption may be valid for certain interconnect geometries and materials, it may not be generally applicable. Hence, Matthiessen's rule needs to be applied with caution.

The work in [8] provides a method to properly integrate the FS and MS models. However, since the conductivity in the FS

model is not spatially resolved, the insights into the effect of integrating grain-boundary and surface scattering model on the spatial resolution of conductivity are lacking.

In this work, we address this gap by building upon the SRFS models [6], [7] and incorporating the grain boundary mechanism based on the original MS theory [8], leading to SRFS-MS model. The SRFS-MS model captures grain boundary scattering through the grain boundary reflectance coefficient ($R$). Instead of merely summing the resistivities associated with the two mechanisms together as per Matthiessen's rule, the SRFS-MS model integrates them by following the method in [8] but appropriately modifying it to account for the spatial resolution. In this manner, the proposed SRFS-MS model maintains the interaction between surface scattering and grain-boundary scattering. At the same time, it offers the spatial resolution of conductivity and explicitly defines its dependence on the specularity of surface scattering ($p$ – which quantifies the degree of elastic scattering from the surface) and $R$ (which quantifies the extent of reflection from the grain boundaries). Furthermore, we propose a circuit compatible version of the SRFS-MS model (utilizing the approach of our SRFS-C3 model), which shows a close match with the physical SRFS-MS model. The key contributions of this work are:

- We develop SRFS-MS model which offers spatial resolution of conductivity and accounts for an integrated effect of surface scattering and grain boundary scattering. We also incorporate temperature ($T$)-dependence in the SRFS-MS model and discuss the effect of $T$ on spatially resolved conductivity.
- We propose an SRFS-MS-C3 model, a circuit-compatible version of the SRFS-MS model based on analytical equations and a few invocations of the physical SRFS-MS model. The SRFS-MS-C3 model obtains the spatially resolved conductivity as a function of $p$, $R$, $T$, $\lambda_0$ and interconnect dimensions.
- We provide insights into how the space-dependence of conductivity is affected by the integrated effect of surface and grain-boundary scattering as well as other physical parameters.

## II. Background

*A. The Mayadas-Shatzkes (MS) theory for Grain Boundary Scattering*

In this sub-section, we provide the relevant details of the MS model that will be utilized for presenting the proposed SRFS-MS model. For more details on the MS approach, we refer the reader to [8].

Computing the electron velocity distribution for scattering by grain boundaries of any shape, size, and orientation is a significant challenge. Therefore, simplifications proposed in [8] are generally used based on physical observations and some assumptions. The first simplification states that for thin films on substrates, the grains are not isotropic but tend to grow in a "columnar" style with the columns perpendicular to the film surfaces (let us define that as the x direction for this subsection as in [8]). The average random distance between two columns is denoted as $d$. Second, the grains extend from the substrate to the top surface of the film, and only the perpendicular grain boundaries are considered. Third, it is assumed that the grain boundaries can be represented by two different types of randomly spaced planes: those parallel to the electric field and those perpendicular to it. The parallel planes produce only elastic reflection (resulting in no energy loss and no increase in resistivity), while the perpendicular planes cause inelastic reflection (leading to an increase in resistivity).

With these approximations, Mayadas and Shatzkes proposed the well-known grain boundary conductivity expression, which utilizes the Boltzmann transport equation in conjunction with the perturbation theory. To account for grain boundary scattering, transition probabilities from one wave vector state $k$ to another $k'$ due to grain boundary scattering are considered. It has been shown that this modifies the relaxation time $\tau$ to $\tau^*$ as follows:

$$\frac{1}{\tau^*} = \frac{1}{\tau} + 2F(|k_x|) \qquad (1)$$

Here the function $F$ is a complex function of $k_x$ (the $x$-component of $k$) as shown below:

$$F(|k_x|) = \frac{\alpha}{2\tau} \frac{k_F}{|k_x|} \frac{1 - exp(-4k_x^2 s^2)}{1 + exp(-4k_x^2 s^2) - 2exp(-k_x^2 s^2)cos(2k_x d)} \qquad (2)$$

$$\alpha = \frac{\lambda_0}{d_{grain}} \left(\frac{R}{1-R}\right) \qquad (3)$$

Here $k_F$ is the Fermi wavevector, $s$ is the standard deviation of the position of the grains (and the assumed delta potential barrier that they introduce), $\lambda_0$ denotes the electron mean free path, $d_{grain}$ represents the grain size, and $R$ is the grain boundary reflectance coefficient.

To simplify (2), the authors in [8] noted that the spacing between the delta potential barriers $d$ can be identified with the average grain diameter $D$ (measured experimentally). This is based on their experimental observation that $k_F^2 s^2 \gg 1$. With this simplification, $F(|k_x|)$ can be written as

$$F(|k_x|) = \frac{\alpha}{2\tau} \frac{k_F}{|k_x|} \qquad (4)$$

where $k_x$ can be written as $k_x = k_F c$ and where $c$ is cosine of the angle that $k_F$ makes with the $x$-axis. The conductivity accounting for the grain boundary as well as the bulk contributions is obtained as

$$\sigma_{GB} = \frac{3}{2} \frac{\sigma_0}{\tau} \int_{-1}^{1} \tau^*(c) c^2 dc \qquad (5a)$$

Substituting (4) in (1) and making an observation that $F(|k_x|)$ and therefore, $\tau^*$ is an even function of $c$, we obtain

$$\sigma_{GB} = 3\sigma_0 \int_0^1 \frac{c^2}{1+\frac{\alpha}{c}} dc \qquad (5b)$$

This can be simplified to

$$\frac{\sigma_{GB}}{\sigma_0} = \left[1 - \frac{3\alpha}{2} + 3\alpha^2 - 3\alpha^3 \ln\left(1 + \frac{1}{\alpha}\right)\right] \qquad (6a)$$

It may be noted that the first term ('1') on the right hand side of (6) represents the bulk conductivity, while the rest of the term accounts for the reduction in conductivity due to grain boundary scattering (defined as $\Delta\sigma_{GB}$). Thus

$$\sigma_{GB} = \sigma_0 - \Delta\sigma_{GB} \qquad (6b)$$

In several works [9], [10], [11], the expression obtained from the MS theory (6) is combined with surface scattering model (such as the FS model) by simply adding the resistivities together according to the Matthiessen's rule. However, based on the MS theory [8], both scattering mechanisms have a deeper connection and need to be integrated together as a whole.

To obtain the overall thin film conductivity ($\sigma_{total}$) accounting for both scattering mechanisms integrated together, we need to follow the same methodology as per the FS theory [5] but replace the relaxation time $\tau$ with $\tau^*$. The final conductivity equation for a thin film of thickness $a$ is shown below:

$$\sigma_{total} = \sigma_{GB} - \frac{6\sigma_0}{\pi \kappa_a}(1-p) \times$$
$$\int_0^{\pi/2} d\Phi \int_1^{\infty} dt \frac{\cos^2 \Phi}{H^2(t, \Phi)} \quad (7)$$
$$\times \left(\frac{1}{t^3} - \frac{1}{t^5}\right) \frac{1 - e^{[-\kappa_a t H(t,\Phi)]}}{1 - p e^{[-\kappa_a t H(t,\Phi)]}}$$

Here, $\Phi$ and $\theta$ are the azimuthal and the polar angles, respectively, $\kappa_a = a/\lambda_0$, $p$ is the surface scattering specularity factor, $t = 1/\cos(\theta)$ and $\sigma_{GB}$ term is obtained from (6). $H(t, \Phi)$ is the ratio of $\tau$ and $\tau^*$ and is obtained as

$$H(t, \Phi) = \frac{\tau}{\tau^*} = 1 + \frac{\alpha}{|\cos(\Phi)|\left(1 - \frac{1}{t^2}\right)^{\frac{1}{2}}} \quad (8)$$

There are two noteworthy points here: (i) In the conductivity equation (7), the first occurrence of $H(t, \Phi)$ is within the argument of the *exp* functions, and the second is in the denominator of the integrand as $H^2(t, \Phi)$ and (ii) the second term on the right hand side of (8) can be interpreted as $\alpha$ divided by the ratio of the x-directed velocity (in the direction of the electron flow) to the magnitude of the velocity ($\frac{v_x}{|v|} = \cos(\Phi)\sin(\theta)$). We will refer back to these points in the subsequent sub-sections.

*B. Spatially Resolved FS (SRFS) Model*

In our previous work [6], we proposed a spatially resolved FS (SRFS) model for rectangular interconnects ($\sigma_{SRFS}$ in equation (9)) derived from the basic Boltzmann Transport Equation. The main advantage of the SRFS model is that conductivity is expressed as a function of the location within the interconnect cross-section (needed for modern interconnect structures, especially vias [6]). Further, as in the original FS model, the SRFS model retains the explicit relationships of conductivity to fundamental electron transport parameters viz. $\lambda_0$ and $p$ and the cross-section width ($a$) and height ($b$). As can be observed from (9), the conductivity is a function of $x_n$ and $y_n$ (the x- and y- locations normalized to $a/2$ and $b/2$, respectively), the bulk copper conductivity ($\sigma_0$), specularity ($p$, where $0 < p < 1$ and $p = 0$ for diffusive and $p = 1$ for completely elastic scattering) and the parameters $\kappa_a (= a/\lambda_0)$ and $\kappa_b (= b/\lambda_0)$. The SRFS model is exact for $p=0$ and approximate for $p>0$. However, the *average* conductivity obtained from the approximate SRFS model shows a reasonably good match with previous works based on FS [12] and Kinetic Theory-based theories [13].

We also proposed the circuit compatible version of the SRFS model in our previous work [7]. With only four sampling points taken from the physical SRFS model, it uses an analytical function to predict the spatial profile with a good accuracy (<0.7%).

With the understanding of the MS and SRFS approaches, let us now present our SRFS-MS model, in which we integrate the MS model within the SRFS equations.

### III. SRFS-MS Model for Rectangular Wires

To incorporate the MS theory into the SRFS model (which is based on FS theory), we account for grain boundary scattering by substituting the relaxation time $\tau$ with $\tau^*$. Recall, the ratio $\tau/\tau^*$ is defined as $H$. As noted before, the $H$ function depends on the component of a unit vector along the direction of current flow (x-direction in Section II A). In the SRFS model for rectangular interconnects, we choose $z$ as the electron flow direction (for ease of accounting for surface scattering in 2 dimensions). Thus, the $H$ function (expressed in terms of a unit vector along the z-direction) is obtained as:

$$H(\theta) = \frac{\tau}{\tau^*} = 1 + \frac{\alpha}{|\cos(\theta)|} \quad (10)$$

Following the procedure for the SRFS model in [6], we first consider $p=0$ and obtain an exact expression for the conductivity. We then derive an approximate expression for conductivity for a general $p$.

*A. Purely Diffusive Surface Scattering (p=0): Exact Solution*

To utilize the Boltzmann Transport Equations (BTE) and integrate the effect of $H(\theta)$, we first obtain the expression for the deviation of the electron distribution function from its equilibrium value ($\Delta f$) similar to [12]. Note, $\Delta f$ is a function of $\tau$; here, we use $\tau^* = \frac{\tau}{H(\theta)}$ instead of $\tau$, and obtain the following expression in terms of the $x$, $y$ and $z$ components of the electron velocity ($v_x$, $v_y$ and $v_z$), the z-directed electric field ($E_z$ i.e. in the current transport direction)

$$\frac{\sigma_{SRFS}(x_n, y_n)}{\sigma_0} = \frac{3}{4\pi} \int_{\theta=0}^{\pi} \eta(x_n, y_n, \theta) * \cos^2\theta \sin\theta d\theta$$

$$\eta(x_n, y_n, \theta) = 2\pi - (1-p)$$
$$\times \left[ 2 \int_{\Phi=0}^{\frac{\pi}{2}} \left( \frac{\left(e^{\frac{-\kappa_a}{2\sin\theta\cos\Phi}} \times \cosh\left(\frac{\kappa_a \times x_n}{2\sin\theta\cos\Phi}\right)\right)}{1 - p e^{\frac{-\kappa_a}{\sin\theta\cos\Phi}}} + \frac{\left(e^{\frac{-\kappa_b}{2\sin\theta\sin\Phi}} \times \cosh\left(\frac{\kappa_b \times y_n}{2\sin\theta\sin\Phi}\right)\right)}{1 - p e^{\frac{-\kappa_b}{\sin\theta\sin\Phi}}} \right) d\Phi \right.$$
$$\left. + \frac{1}{2} \sum_{n,d} \int_{\Phi=0}^{\frac{\pi}{2}} \left| \frac{e^{\frac{-\kappa_a \times d}{2\sin\theta\cos\Phi}}}{1 - p e^{\frac{-\kappa_a}{\sin\theta\cos\Phi}}} - \frac{e^{\frac{-\kappa_b \times n}{2\sin\theta\sin\Phi}}}{1 - p e^{\frac{-\kappa_b}{\sin\theta\sin\Phi}}} \right| d\Phi \right]$$

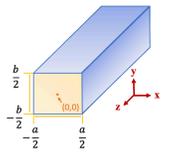

Where $(n,d) \to \begin{Bmatrix} (1+y_n, 1+x_n), (1-y_n, 1+x_n), \\ (1+y_n, 1-x_n), (1-y_n, 1-x_n) \end{Bmatrix}$  $x_n = \frac{x}{a/2} \in [-1,1]$, $y_n = \frac{y}{b/2} \in [-1,1]$;  $\kappa_a = \frac{a}{\lambda_0}$, $\kappa_b = \frac{b}{\lambda_0}$  (9)

$$\Delta f_{s1,s2} = \frac{q\tau E_z}{H(\theta)m_{eff}}\frac{\partial f_0}{\partial v_z}\left(1 - \exp\left(-\frac{H(\theta)\min\left(\frac{x+(s1)\frac{a}{2}}{v_x}, \frac{y+(s2)\frac{b}{2}}{v_y}\right)}{\tau}\right)\right) \quad (11)$$

Here, s1=sign($v_x$), s2=sign($v_y$), i.e. $s1, s2 \in \{-1, +1\}$, leading to four possible regimes for $\Delta f$ (i.e. $\Delta f_{+1,+1}$, $\Delta f_{-1,+1}$, $\Delta f_{+1,-1}$, and $\Delta f_{-1,-1}$). Other parameters used in (11) include $q$ (electronic charge), $m_{eff}$ (effective mass of the electrons) and $f_0$ (the equilibrium distribution function of the electrons). Using the $\Delta f$ functions of (11) in the Boltzmann Transport Equation (BTE), we obtain the current density (in the $z$ direction) as

$$J_z = -2q\left(\frac{m_{eff}}{h}\right)^3 \int v_z f d\vec{v} =$$
$$= -\frac{2q^2 m_{eff}^2 E_z}{h^3}\sum_{s1,s2}\int_0^\infty\int_{\theta=0}^\pi\int_{\Phi=\Phi_l}^{\Phi_h}\frac{\tau}{H(\theta)}\frac{\partial f_0}{\partial v_z}v_z v^2 \sin\theta$$
$$\times\left(1 - \exp\left(-\frac{H(\theta)\min\left(\frac{x+(s1)\frac{a}{2}}{v_x}, \frac{y+(s2)\frac{b}{2}}{v_y}\right)}{\tau}\right)\right)dvd\Phi d\theta \quad (12)$$

Here, $v$ refers to the magnitude of the velocity and $(\Phi_l, \Phi_h)$=$(0,\frac{\pi}{2})$, $(\frac{\pi}{2},\pi)$, $(\pi,\frac{3\pi}{2})$, $(\frac{3\pi}{2},2\pi)$ for (s1,s2)=(+1,+1), (-1,+1), (-1,-1), (+1,-1), respectively. Now, using $\frac{\partial f_0}{\partial v_z} = \frac{\partial f_0}{\partial v}\frac{v_z}{v}$, applying the degenerate electron gas assumption ($\int_0^\infty g(v)\frac{\partial f_0}{\partial v}dv = -g(\tilde{v})$ where $\tilde{v}$ is the electron velocity at the Fermi surface- see [5], [6] for more details), and utilizing $\lambda_0 = \tau\tilde{v}$, the spatially-resolved conductivity $\sigma(x,y) = \frac{J_z}{E_z}$ is obtained as

$$\sigma(x,y) = -\frac{2q^2 m_{eff}^2}{h^3}\sum_{s1,s2}\int_{\theta=0}^\pi\int_{\Phi=\Phi_l}^{\Phi_h}\frac{\lambda_0 \tilde{v}^2}{H(\theta)}\cos^2\theta\sin\theta$$
$$\times\left(1 - \exp\left(-H(\theta)\min\left(\frac{x+(s1)\frac{a}{2}}{\lambda_0\sin\theta\cos\Phi}, \frac{y+(s2)\frac{b}{2}}{\lambda_0\sin\theta\sin\Phi}\right)\right)\right)d\Phi d\theta \quad (13)$$

Note, the function $H(\theta)$ appears in the denominator of the integrand and as arguments of the exponential function, which will be discussed subsequently. Also, note that if our interest is in the conductivity averaged over the cross-section, we need to integrate (12) with respect to $x$ and $y$. This integration yields an additional $H(\theta)$ term in the denominator (since the variable of integration is multiplied by this term). This explains the $H^2$ term in the derivation in [8] (see previous section). Since, here, we are interested in the spatially resolved conductivity (not the average), we do not obtain that additional $H(\theta)$ term and hence, have the first power of $H$ in the denominator.

Now, we express (13) in terms of $\sigma_0\left(=\frac{8\pi}{3}\frac{q^2 m_{eff}^2 \tilde{v}^2}{h^3}\lambda_0\right)$, normalized dimensions $x_n = \frac{x}{a/2}$ and $y_n = \frac{y}{b/2}$ and parameters $\kappa_a(=a/\lambda_0)$ and $\kappa_b(=b/\lambda_0)$ to obtain

$$\sigma(x_n, y_n) = \frac{3}{4\pi}\sigma_0\sum_{s1,s2}\int_{\theta=0}^\pi\int_{\Phi=\Phi_l}^{\Phi_h}\frac{1}{H(\theta)}\cos^2\theta\sin\theta$$
$$\times\left(1 - \exp\left(-H(\theta)\min\left(\frac{\kappa_a(x_n+s1)}{2\sin\theta\cos\Phi}, \frac{\kappa_b(y_n+s2)}{2\sin\theta\sin\Phi}\right)\right)\right)d\Phi d\theta \quad (14)$$

Following the procedure in [6], we obtain the integral in (14) with respect to $\Phi$ and calculate the sum for all $s1$ and $s2$. We define the resultant as $\eta(x_n, y_n, \theta)$. Also, since $H(\theta)$ is a function of $|\cos(\theta)|$, the integral in (14) with respect to $\theta$ has the same result when integrated from 0 to π/2 and π/2 to π. Thus,

$$\sigma(x_n, y_n) = \frac{3}{2\pi}\sigma_0\int_{\theta=0}^{\pi/2}\frac{\eta(x_n, y_n, \theta)}{H(\theta)}\cos^2\theta\sin\theta\, d\theta \quad (15)$$

For θ between 0 and π/2, $|\cos(\theta)|$ can be replaced by $\cos(\theta)$ in $H(\theta)$. Thus,

$$H(\theta) = \frac{\tau}{\tau^*} = 1 + \frac{\alpha}{\cos(\theta)}, \text{ for } \theta \in [0, \pi/2] \quad (16)$$

To obtain $\eta$, we change the limits of integration from $(\Phi_l, \Phi_h)$ to $(0, \frac{\pi}{2})$ for all $s1$ and $s2$, which gives

$$\eta(x_n, y_n, \theta) = \sum_{s1,s2}\int_{\Phi=0}^{\frac{\pi}{2}}\left(1 - \exp\left(-H(\theta)\min\left(\frac{\kappa_a(1+(s1)x_n)}{2\sin\theta\cos\Phi}, \frac{\kappa_b(1+(s2)y_n)}{2\sin\theta\sin\Phi}\right)\right)\right)d\Phi \quad (17)$$

Further, we observe that the min function on the right hand side of (17) implies that the argument of the exp function must use the term with $x_n$ if $\tan(\Phi) < \frac{b}{a}\left(\frac{1+(s2)y_n}{1+(s1)x_n}\right)$; else, it uses the term with $y_n$. Thus,

$$\eta(x_n, y_n, \theta) = \sum_{s1,s2}\int_0^{\tan^{-1}\left(\frac{b}{a}\left(\frac{1+(s2)y_n}{1+(s1)x_n}\right)\right)}\left(1 - \exp\left(-H(\theta)\frac{\kappa_a(1+(s1)x_n)}{2\sin\theta\cos\Phi}\right)\right)d\Phi$$
$$+ \int_{\tan^{-1}\left(\frac{b}{a}\left(\frac{1+(s2)y_n}{1+(s1)x_n}\right)\right)}^{\pi/2}\left(1 - \exp\left(-H(\theta)\frac{\kappa_a(1+(s2)y_n)}{2\sin\theta\sin\Phi}\right)\right)d\Phi \quad (18)$$

Defining $(n, d)$ as $\{(1+y_n, 1+x_n), (1-y_n, 1+x_n), (1+y_n, 1-x_n), (1-y_n, 1-x_n)\}$ for the four cases corresponding to different values of $s1$ and $s2$, we obtain

$$\eta(x_n, y_n, \theta) = \sum_{n,d}\int_0^{\tan^{-1}\left(\frac{b}{a}\frac{n}{d}\right)}\left(1 - \exp\left(-H(\theta)\frac{\kappa_a d}{2\sin\theta\cos\Phi}\right)\right)d\Phi$$
$$+ \int_{\tan^{-1}\left(\frac{b}{a}\frac{n}{d}\right)}^{\pi/2}\left(1 - \exp\left(-H(\theta)\frac{\kappa_b n}{2\sin\theta\sin\Phi}\right)\right)d\Phi \quad (19)$$

It is noteworthy that $(n, d)$ defined as the boundary points in [6] do not change due to the added term $H(\theta)$. Hence the same procedure as given in the Appendix of [6] can be followed to simplify (17) to obtain the following expressions for $\eta$.

$$\eta(x_n, y_n, \theta) = 2\pi - 2\int_0^{\pi/2}\left\{e^{\frac{-H(\theta)\kappa_b}{2\sin\theta\sin\Phi}}\cosh\left(\frac{H(\theta)\kappa_b y_n}{2\sin\theta\sin\Phi}\right)\right.$$
$$\left. + e^{\frac{-H(\theta)\kappa_a}{2\sin\theta\cos\Phi}}\cosh\left(\frac{H(\theta)\kappa_a x_n}{2\sin\theta\cos\Phi}\right)\right\}d\Phi$$
$$-\frac{1}{2}\sum_{n,d}\int_0^{\pi/2}\left|e^{\frac{-H(\theta)\kappa_a d}{2\sin\theta\cos\Phi}} - e^{\frac{-H(\theta)\kappa_b n}{2\sin\theta\sin\Phi}}\right|d\Phi \quad (10)$$

where $(n, d) \to \{(1+y_n, 1+x_n), (1-y_n, 1+x_n), (1+y_n, 1-x_n), (1-y_n, 1-x_n)\}$

This can be expressed as

$$\eta(x_n, y_n, \theta) = 2\pi - \eta_{SS\leftarrow GB}(x_n, y_n, \theta) \quad (11)$$

where $\eta_{SS \leftarrow GB}(x_n, y_n, \theta)$ comprises the two integrals on the right hand side of (18) and accounts for the reduction in conductivity due to the surface scattering (with the integrated effect of grain boundary scattering included via $H(\theta)$).

When (20) is substituted in (15), we get two terms – one corresponding to $2\pi$ and the other due to $\eta_{SS \leftarrow GB}(x_n, y_n, \theta)$. The first term (which is spatially uniform and is defined as $\sigma_1$ for now) yields

$$\sigma_1 = \frac{3}{2\pi}\sigma_0 \int_{\theta=0}^{\pi/2} \frac{2\pi}{H(\theta)} cos^2\theta sin\theta\, d\theta = 3\sigma_0 \int_{\theta=0}^{\pi/2} \frac{cos^2\theta sin\theta}{1 + \alpha/cos\theta} d\theta \quad (12)$$

When simplified, (22) yields the integral in equation (5b) and therefore, $\sigma_1 = \sigma_{GB} = \sigma_0 - \Delta\sigma_{GB}$ as given by (6).

The second term in (21), when substituted in (15) yields the spatially resolved conductivity reduction due to integrated effect of surface scattering and grain boundary scattering (defined as $\Delta\sigma_{SS \leftarrow GB}$). Putting everything together, the final conductivity model for $p=0$ is given by

$$\sigma(x_n, y_n) = \sigma_0 - \Delta\sigma_{GB} - \Delta\sigma_{SS \leftarrow GB}(x_n, y_n)$$

where

$$\Delta\sigma_{GB} = \sigma_0\left[\frac{3\alpha}{2} - 3\alpha^2 + 3\alpha^3 \ln\left(1 + \frac{1}{\alpha}\right)\right]$$

$$\Delta\sigma_{SS \leftarrow GB}(x_n, y_n) = \frac{3}{2\pi}\sigma_0 \int_{\theta=0}^{\pi/2} \frac{\eta_{SS \leftarrow GB}(x_n, y_n, \theta)}{H(\theta)} cos^2\theta sin\theta\, d\theta$$

$$\eta_{SS \leftarrow GB}(x_n, y_n, \theta) = 2\int_0^{\pi/2} \left\{ e^{\frac{-H(\theta)\kappa_b}{2sin\theta sin\Phi}} cosh\left(\frac{H(\theta)\kappa_b y_n}{2sin\theta sin\Phi}\right) \right.$$
$$\left. + e^{\frac{-H(\theta)\kappa_a}{2sin\theta cos\Phi}} cosh\left(\frac{H(\theta)\kappa_a x_n}{2sin\theta cos\Phi}\right) \right\} d\Phi \quad (13)$$
$$+ \frac{1}{2}\sum_{n,d}\int_0^{\pi/2} \left|e^{\frac{-H(\theta)\kappa_a d}{2sin\theta cos\Phi}} - e^{\frac{-H(\theta)\kappa_b n}{2sin\theta sin\Phi}}\right| d\Phi$$

$(n,d) \to \{(1+y_n, 1+x_n), (1-y_n, 1+x_n), (1+y_n, 1-x_n), (1-y_n, 1-x_n)\}$

*B. Specular Scattering ($0 < p \leq 1$): Approximate Solution*

For general specularity, it is challenging to obtain an exact solution for a rectangular wire; therefore, we adopt the approach presented in [6], which approximates the $\Delta f$ functions such that (i) they converge to the exact solution for $p=0$ (derived in the previous sub-section) and trivial case for $p=1$, (ii) they converge to the exact solution for general $p$ for thin films [8] and (iii) the method proposed in [6] to simplify (19) to (20) is applicable. This approximation essentially considers interaction between surface scattering events from parallel sidewalls only. While better approximations may be possible with further investigation, we pursue with this approach as it shows a reasonably good agreement with the trends from previous average conductivity models [6], [13]. Hence, $\Delta f_{s1,s2}$ is obtained as

$$\Delta f_{s1,s2} = \frac{q\tau E_z}{H(\theta)m_{eff}} \frac{\partial f_0}{\partial v_z} \times$$
$$\left(1 - (1-p) \times \max\left\{\frac{e^{-\frac{H(\theta)(x+(s1)\frac{a}{2})}{\tau v_x}}}{1 - p\, e^{-aH(\theta)/\tau v_x}}, \frac{e^{-\frac{H(\theta)(y+(s2)\frac{b}{2})}{\tau v_y}}}{1 - p\, e^{-aH(\theta)/\tau v_y}}\right\}\right) \quad (14)$$

Now, following the same process as for $p=0$ in the previous sub-section, we obtain the final conductivity model as shown at the bottom of this page as equation (25). The key features of the proposed SRFS-MS model are summarized below:

- The model is based on conductivity subtraction (more physical than resistivity addition).
- The model directly relates conductivity to physical parameters viz. $\sigma_0, \lambda_0, p, R, d_{grain}$ and the wire cross-section geometry ($a$ and $b$).
- The model goes beyond the simple Matthessian's rule and integrates the effect of grain-boundary and surface scattering.
- The model offers spatial resolution of conductivity due to surface scattering (integrated with grain boundary scattering).
- This model converges to a thin film of thickness $a$ if $\kappa_b \to \infty$.
- If $H(\theta)$ is 1, the model reduces to the Matthessian's approach of combining various components.
- Since the model is an explicit function of physical parameters, it is seamless to incorporate temperature ($T$)-dependence (which is also shown in (25) and

$$\sigma_{SRFS-MS}(x_n, y_n, T) = \sigma_0(T) - \Delta\sigma_{GB}(T) - \Delta\sigma_{SS \leftarrow GB}(x_n, y_n, T)$$

$$\Delta\sigma_{GB}(T) = \sigma_0(T)\left[\frac{3\alpha(T)}{2} - 3\alpha(T)^2 + 3\alpha(T)^3 \ln\left(1 + \frac{1}{\alpha(T)}\right)\right]$$

$$\sigma_{SS \leftarrow GB}(x_n, y_n, T) = \frac{3}{2\pi}\sigma_0(T)\int_{\theta=0}^{\frac{\pi}{2}} \eta_{SS \leftarrow GB}(x_n, y_n, \theta, T) \frac{cos^2\theta}{H(\theta, T)} sin\theta d\theta$$

$$\eta_{SS \leftarrow GB}(x_n, y_n, \theta, T) = (1-p)$$
$$\times \left[ 2\int_{\Phi=0}^{\frac{\pi}{2}} \left(\frac{\left(e^{\frac{-\kappa_a(T) \times H(\theta,T)}{2sin\theta cos\Phi}} \times cosh\left(\frac{\kappa_a(T) \times x_n \times H(\theta,T)}{2sin\theta cos\Phi}\right)\right)}{1 - pe^{\frac{-\kappa_a(T) \times H(\theta,T)}{sin\theta cos\Phi}}} + \frac{\left(e^{\frac{-\kappa_b(T) \times H(\theta,T)}{2sin\theta sin\Phi}} \times cosh\left(\frac{\kappa_b(T) \times y_n \times H(\theta,T)}{2sin\theta sin\Phi}\right)\right)}{1 - p*e^{\frac{-\kappa_b(T) \times H(\theta,T)}{sin\theta sin\Phi}}}\right) d\Phi \right.$$
$$\left. + \frac{1}{2}\sum_{n,d}\int_{\Phi=0}^{\frac{\pi}{2}} \left|\frac{e^{\frac{-\kappa_a(T) \times d \times H(\theta,T)}{2sin\theta cos\Phi}}}{1 - pe^{\frac{-\kappa_a(T) \times H(\theta,T)}{sin\theta cos\Phi}}} - \frac{e^{\frac{-\kappa_b(T) \times n \times H(\theta,T)}{2sin\theta sin\Phi}}}{1 - pe^{\frac{-\kappa_b(T) \times H(\theta,T)}{sin\theta sin\Phi}}}\right| d\Phi \right]$$

$$H(\theta, T) = 1 + \frac{\alpha(T)}{cos(\theta)} \qquad \alpha(T) = \frac{\lambda_0(T)}{d_{grain}}\left(\frac{R}{1-R}\right)$$

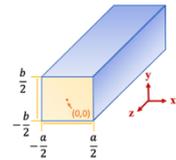

$(n,d) \to \begin{Bmatrix} (1+y_n, 1+x_n), (1-y_n, 1+x_n), \\ (1+y_n, 1-x_n), (1-y_n, 1-x_n) \end{Bmatrix} \quad x_n = \frac{x}{a/2} \in [-1,1],\ y_n = \frac{y}{b/2} \in [-1,1];\quad \kappa_a(T) = \frac{a}{\lambda_0(T)}, \kappa_b(T) = \frac{b}{\lambda_0(T)}$

(25)

discussed next).

*C. Incorporating Temperature Dependence in the SRFS-MS Model*

For temperature dependence, we utilize the fact that $\frac{\sigma_0}{\lambda_0}$ is constant for a material (from electron gas theory). Let us define it as $c_0$. Thus,

$$c_0 = \frac{\sigma_0}{\lambda_0} = \frac{8\pi}{3} \frac{q^2 m_{eff}^2 \tilde{v}^2}{h^3} \quad (26)$$

Then, we use a well-known expression for $\sigma_0(T)$ [12]

$$\frac{1}{\sigma_0(T)} = \frac{1}{c_0 \lambda_r} + \frac{6\pi^2 c_L}{c_0^2} \frac{q^2}{h\tilde{v}^2} \frac{1}{k_B \Theta} \left(\frac{T}{\Theta}\right)^5 \int_0^{\Theta/T} \frac{z^5}{(e^z-1)(1-e^{-z})} dz \quad (27)$$

Here, $\lambda_r$ is the residual mean free path (significant at very low temperatures), $c_L$ is a lattice-dependent constant (inversely proportional to the mass of the atom and volume of the unit cell, and captures the interactions between electrons and lattice), $k_B$ is the Boltzmann constant and $\Theta$ is the Debye temperature.

Further, from (26) and (27), we obtain

$$\lambda_0(T) = \sigma_0(T)/c_0 = \lambda_{0\_RT} \sigma_0(T)/\sigma_{0\_RT} \quad (28)$$

Here, $\sigma_{0\_RT}$ and $\lambda_{0\_RT}$ are the room temperature values of $\sigma_0$ and $\lambda_0$, respectively.

With these temperature dependencies of $\sigma_0$ and $\lambda_0$, we observe that in the SRFS-MS model, $H, \alpha, \kappa_a, \kappa_b$ also become temperature-dependent, as shown in (25).

## IV. SRFS-MS CIRCUIT-COMPATIBLE CONDUCTIVITY (SRFS-MS-C3) MODEL

We now propose the circuit compatible version of the SRFS-MS model with the motivation to reduce the computation of the complex integrals of (25) and use analytical functions to predict the spatial profile of the conductivity due to the integrated effect of surface scattering and grain boundary scattering. We also incorporate the temperature-dependent effects in this model.

We follow the approach presented in [7] for the SRFS-C3 model. The SRFS-MS-C3 model obtains conductivity values for four points in the rectangular cross-section (three points for a symmetric square cross-section) from the SRFS-MS model of (25) and predicts the conductivity for the entire cross-section of the wire using a similar analytical function as in [7] (with some differences which we discuss later in this section). The four points obtained from SRFS-MS model are at (i) $x_n=0$, $y_n=0$, (ii) $x_n=1$, $y_n=0$, (iii) $x_n=0$, $y_n=1$ and (iv) $x_n=1$, $y_n=1$. The analytical function for the x-direction using in SRFS-MS-C3 is given below (the empirical rationale for using this function can be found in [7])

$$\begin{aligned}\xi(x_n,\beta_a) =& (1+x_n)\{cosh(\beta_a)[Chi(\beta_a(1+x_n))-Chi(\beta_a)] \\ & - sinh(\beta_a)[Shi(\beta_a(1+x_n))-Shi(\beta_a)]\} \\ & +(1-x_n)\{cosh(\beta_a)[Chi(\beta_a(1-x_n))-Chi(\beta_a)] \\ & - sinh(\beta_a)[Shi(\beta_a(1-x_n))-Shi(\beta_a)]\}\end{aligned} \quad (29)$$

Here, $\beta_a$ is the model parameter, which is a function of $a$ (or $\kappa_a$) and other physical parameters (discussed, in detail subsequently). *Chi* and *Shi* are the cosh integral and sinh integral functions. We use the same function for the y-direction with $x_n$ replaced by $y_n$ and $\beta_a$ replaced by $\beta_b$, where $\beta_b$ has a similar form as $\beta_a$ except that $\kappa_a$ is replaced by $\kappa_b$.

In the SRFS-C3 model [7] (in which the objective is to model the surface scattering only), $\beta_a(\beta_b)$ are functions of $p$ and $\kappa_a$ ($\kappa_b$). In the proposed SRFS-MS-C3 model, $\beta_a(\beta_b)$ must be is a function of $p$, $\kappa_a$ ($\kappa_b$) and $R$. To obtain the new $\beta_a(\beta_b)$ function, we draw the attention of the readers to two observations: (i) $\xi$ function models the *shape* of the spatial profile of conductivity and therefore, is mainly, used to capture the average effect of $x_n$- and $y_n$-dependent functions in $\eta_{SS\leftarrow GB}$ (see (25)). The absolute values are mainly obtained from the four sample points of the SRFS-MS model. (ii) The difference between the SRFS-C3 and SRFS-MS-C3 is mainly the incorporation of the effect of $H(\theta)$ function in the model. Therefore, let us first understand the effect of $H(\theta)$ on the $x_n$- and $y_n$- dependent functions in (25).

From (25), we can observe that $H(\theta)$ is multiplied by $\kappa_a$ ($=a/\lambda_0$) and $\kappa_b$($=b/\lambda_0$) terms. One can think of this leading to effective values of $\kappa_a$ and $\kappa_b$ (defined as $\widehat{\kappa_a}$ and $\widehat{\kappa_b}$). Thus

$$\widehat{\kappa_a} = \kappa_a \langle H(\theta) \rangle = \kappa_a \left(1 + \alpha \left\langle \frac{1}{cos(\theta)} \right\rangle\right) \quad (30)$$

Here, $\langle . \rangle$ refers to an "average" effect due to the integration with respect to $\theta$. Now, substituting (3) for $\alpha$ in (20) and using $\kappa_a$ ($=a/\lambda_0$), we obtain

$$\begin{aligned}\widehat{\kappa_a} &= \kappa_a + \kappa_a \left\langle \frac{1}{cos(\theta)} \right\rangle \frac{\lambda_0}{d_{grain}} \left(\frac{R}{1-R}\right) \\ &= \kappa_a + \left\langle \frac{1}{cos(\theta)} \right\rangle \frac{a}{d_{grain}} \left(\frac{R}{1-R}\right)\end{aligned} \quad (31)$$

Typically, (and especially for scaled interconnects), $d_{grain}$ is a function of $a$ and $b$ (see [14], for instance); however, we approximate $\left(\left\langle \frac{1}{cos(\theta)} \right\rangle \frac{a}{d_{grain}}\right)$ as a constant ($K$), neglecting the dependence of this factor on $a$ and $b$ for simplicity of the circuit-compatible model. Thus,

$$\widehat{\kappa_a} = \kappa_a + K \left(\frac{R}{1-R}\right) \quad (32)$$

Similarly,

$$\widehat{\kappa_b} = \kappa_b + K \left(\frac{R}{1-R}\right) \quad (33)$$

Now, we obtain model fitting parameter $\beta_a$ by minimizing the mean square error (MSE) between the SRFS-MS and SRFS-MS-C3 expressions. As in the SRFS-C3 model, we maintain model simplicity by fitting $\beta_a$ and $\beta_b$ independently, thereby neglecting the effect of cross interactions between $x$ and $y$ on the estimation of $\beta_a$ and $\beta_b$. This also enable us to deduce the analytical expressions with minimal dependence on the aspect

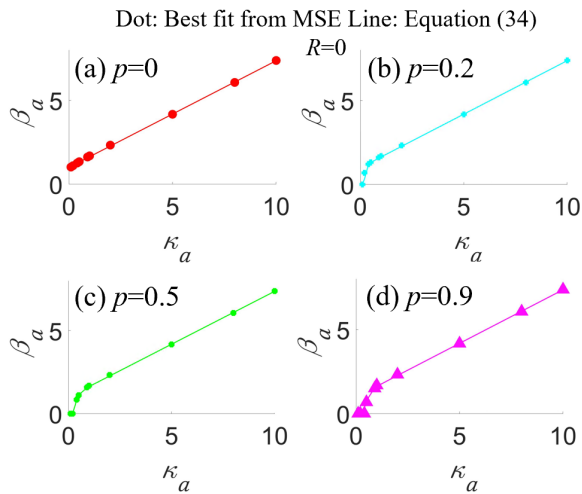

Fig. 1. $\beta_a$ versus $\kappa_a$ for $R=0$ showing a close match between Equation (34) and the best fit from MSE for a) $p=0$ b) $p=0.2$ c) $p=0.5$ d) $p=0.9$

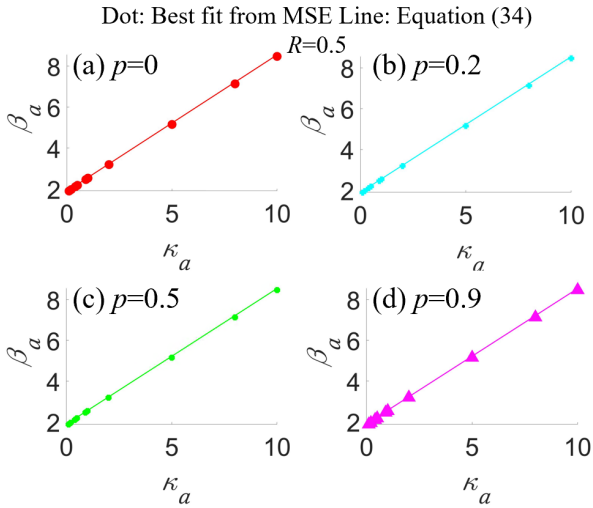

Fig. 2. $\beta_a$ versus $\kappa_a$ for $R=0.5$ showing a close match between Equation (34) and the best fit from MSE for a) $p=0$ b) $p=0.2$ c) $p=0.5$ d) $p=0.9$

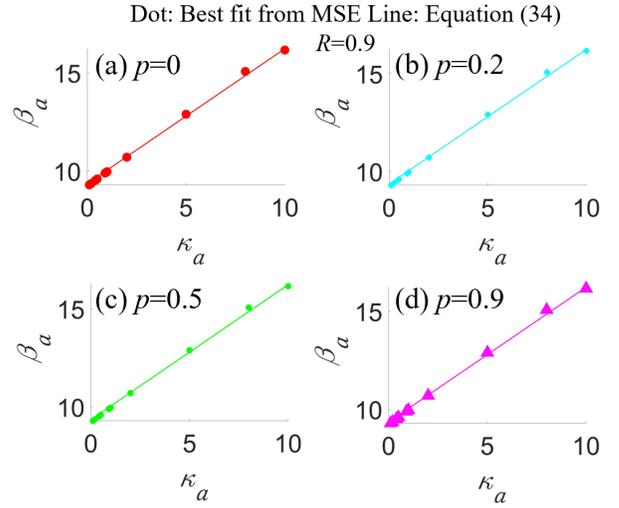

Fig. 3. $\beta_a$ versus $\kappa_a$ for $R=0.9$ showing a close match between Equation (34) and the best fit from MSE for a) $p=0$ b) $p=0.2$ c) $p=0.5$ d) $p=0.9$

ratio ($a/b$) of the wire cross-section, thereby averting more complexities with insignificant loss in accuracy (details later). We repeat this process for various values of $\kappa_a$ ($\kappa_b$), $p$ and $R$, and obtain the trends of $\beta_a$ ($\beta_b$) as a function of these physical parameters. We then empirically deduce an analytical equation for $\beta_a$ ($\beta_b$) based on these trends.

The outcomes of this process are shown in Fig. 1-3. Our discussion will be presented in terms of $\beta_a$ and $\kappa_a$. The same is applicable to $\beta_b$ (replacing $\beta_a$ in the expression below) and $\kappa_b$ (replacing $\kappa_a$). The optimal $\beta_a$ values (obtained by minimizing the MSE between SRFS-MS and SRFS-MS-C3) for various combinations of $\kappa_a$, $p$ and $R$ are shown in Fig 1-3 as markers. The relationship of $\beta_a$ with $\kappa_a$, $p$ and $R$ can be approximately captured using the following equation:

$$\beta_a = max\left[\widehat{m}\widehat{\kappa_a} + 1 + ln\left(tanh\frac{\widehat{\kappa_a}^2}{\varepsilon p}\right), 0.01\right] \quad (34)$$

Note, here we use $\widehat{\kappa_a}$, as derived in (32). Further, $\varepsilon$ is the fitting parameter inherited from the SRFS-C3 model [4] and is equal to 0.42. $K$ in (32) is an additional fitting parameter and is found to be equal to 1.35 by minimizing the MSE between (34) and the best fit values of $\beta_a$. For $\widehat{m}$, we notice that the slope of $\beta_a$ with respect to $\kappa_a$ is $2/\pi$ for $R=0$ (as in SRFS-C3) and $ln(2)$ for $R=1$, and changes ~linearly between the two extreme values of $R$. Hence, we obtain

$$\widehat{m} = \frac{2}{\pi} + R \times \left(ln(2) - \frac{2}{\pi}\right) \quad (35)$$

The complete SRFS-MS-C3 model is summarized at the bottom of last page (36). Note, in the SRFS-MS-C3 model, the four sample points $\sigma_{SRFS-MS}(0,0)$, $\sigma_{SRFS-MS}(0,1)$, $\sigma_{SRFS-MS}(1,0)$ and $\sigma_{SRFS-MS}(1,1)$ can be replaced by the corresponding conductivity subtraction formulae from (25) (i.e. $\sigma_{SRFS-MS} = \sigma_0 - \Delta\sigma_{GB} - \Delta\sigma_{SS\leftarrow GB}$). Since the only space-dependent terms correspond to $\Delta\sigma_{SS\leftarrow GB}$, one can replace $\sigma_{SRFS-MS}$ with $\Delta\sigma_{SS\leftarrow GB}$ in the terms multiplied with the $\xi$ functions. Thus, one

$$\sigma_{SRFS-MS-C3}(x_n, y_n, T)$$
$$= \sigma_{SRFS-MS}(0,0,T) - \left(\sigma_{SRFS-MS}(0,0,T) - \sigma_{SRFS-MS}(1,0,T)\right) \times \frac{\xi(x_n, \beta_a, T)}{\xi_1(\beta_a, T)}$$
$$- \left(\sigma_{SRFS-MS}(0,0,T) - \sigma_{SRFS-MS}(0,1,T)\right) \times \frac{\xi(y_n, \beta_b, T)}{\xi_1(\beta_b, T)}$$
$$-\left(\sigma_{SRFS-MS}(0,1,T) + \sigma_{SRFS-MS}(1,0,T) - \sigma_{SRFS-MS}(0,0,T) - \sigma_{SRFS-MS}(1,1,T)\right) \times \frac{\xi(x_n, \beta_a, T)}{\xi_1(\beta_a, T)} \frac{\xi(y_n, \beta_b, T)}{\xi_1(\beta_b, T)}$$
$$\xi(x_n, \beta_a, T) = \xi(x_n, \beta_a(T)) = (1+x_n)\{cosh(\beta_a)[Chi(\beta_a(1+x_n)) - Chi(\beta_a)] - sinh(\beta_a)[Shi(\beta_a(1+x_n)) - Shi(\beta_a)]\}$$
$$+(1-x_n)\{cosh(\beta_a)[Chi(\beta_a(1-x_n)) - Chi(\beta_a)] - sinh(\beta_a)[Shi(\beta_a(1-x_n)) - Shi(\beta_a)]\}$$
$$\xi(y_n, \beta_b, T) = \xi(y_n, \beta_b(T)) = (1+y_n)\{cosh(\beta_b)[Chi(\beta_b(1+y_n)) - Chi(\beta_b)] - sinh(\beta_b)[Shi(\beta_b(1+y_n)) - Shi(\beta_b)]\}$$
$$+(1-y_n)\{cosh(\beta_b)[Chi(\beta_b(1-y_n)) - Chi(\beta_b)] - sinh(\beta_b)[Shi(\beta_b(1-y_n)) - Shi(\beta_b)]\}$$
$$\xi_1(\beta_a, T) = \xi_1(\beta_a(T)) = 2 \times [cosh(\beta_a)(Chi(2\beta_a) - Chi(\beta_a)) - sinh(\beta_a)(Shi(2\beta_a) - Shi(\beta_a))]$$
$$\xi_1(\beta_b, T) = \xi_1(\beta_b(T)) = 2 \times [cosh(\beta_b)(Chi(2\beta_b) - Chi(\beta_b)) - sinh(\beta_b)(Shi(2\beta_b) - Shi(\beta_b))]$$
$$\beta_a(T) = max\left[\widehat{m} \times \widehat{\kappa_a}(T) + 1 + ln\left(tanh\frac{\widehat{\kappa_a}^2(T)}{0.42p}\right), 0.01\right], \quad \beta_b(T) = max\left[\widehat{m} \times \widehat{\kappa_b}(T) + 1 + ln\left(tanh\frac{\widehat{\kappa_b}^2(T)}{0.42p}\right), 0.01\right]$$
$$\widehat{m} = \frac{2}{\pi} + R \times \left(ln(2) - \frac{2}{\pi}\right), \widehat{\kappa_a}(T) = \kappa_a(T) + \left(\frac{1.35R}{1-R}\right), \widehat{\kappa_b}(T) = \kappa_b(T) + \left(\frac{1.35R}{1-R}\right)$$
$$x_n = \frac{x}{a/2} \in [-1,1], \quad y_n = \frac{y}{b/2} \in [-1,1]; \quad \kappa_a(T) = \frac{a}{\lambda_0(T)}, \kappa_b(T) = \frac{b}{\lambda_0(T)}; \quad (36)$$

can write the SRFS-MS-C3 model based on conductivity subtraction as follows:

$$\begin{aligned}\sigma_{SRFS-MS-C3}&(x_n, y_n, T)\\ &= \sigma_0(T) - \Delta\sigma_{GB}(T) - \Delta\sigma_{SS\leftarrow GB}(0,0,T)\\ &\quad - \big(\Delta\sigma_{SS\leftarrow GB}(0,0,T)\\ &\quad - \Delta\sigma_{SS\leftarrow GB}(1,0,T)\big)\times\frac{\xi(x_n,\beta_a,T)}{\xi_1(\beta_a,T)}\\ &\quad - \big(\Delta\sigma_{SS\leftarrow GB}(0,0,T)\\ &\quad - \Delta\sigma_{SS\leftarrow GB}(0,1,T)\big)\times\frac{\xi(y_n,\beta_b,T)}{\xi_1(\beta_b,T)}\\ &\quad -\big(\Delta\sigma_{SS\leftarrow GB}(0,1,T) + \Delta\sigma_{SS\leftarrow GB}(1,0,T) - \Delta\sigma_{SS\leftarrow GB}(0,0,T)\\ &\quad - \Delta\sigma_{SS\leftarrow GB}(1,1,T)\big)\times\frac{\xi(x_n,\beta_a,T)}{\xi_1(\beta_a,T)}\frac{\xi(y_n,\beta_b,T)}{\xi_1(\beta_b,T)}\end{aligned} \quad (37)$$

Thus, the SRFS-MS-C3 model retains the conductivity subtraction attribute of SRFS-MS to account for various scattering mechanisms which is also suggested by the physical FS and MS theories [7].

The summarized SRFS-MS-C3 model in (36) also shows the incorporation of temperature dependence. This is achieved in two ways: (i) The four sample points obtained from SRFS-MS models already have the temperature-dependence (see (25)) in their absolute values and (ii) the *shape* of the spatial profile of conductivity modeled using $\xi$ function incorporates $T$-dependence in $\beta_a$ and $\beta_b$ via $\kappa_a(T) = a/\lambda_0(T)$ and $\kappa_b = b/\lambda_0(T)$ using the equation (28).

## V. SEMI-ANALYTICAL AVERAGE CONDUCTIVITY MODEL DERIVED FROM SRFS-MS-C3

By virtue of the analytical equations of the SRFS-MS-C3 and the fact that it can be expressed in terms of conductivity terms subtracted (rather than resistivity terms added), one can integrate (36) with respect to $x$ and $y$ (and divide by $ab$) to obtain the semi-analytical expression for the average conductivity. This is presented in (38). Note, the expression uses $\beta_a$ and $\beta_b$, which are explicit functions of physical parameters $p$, $R$, $a$, $b$, $\lambda_0$ and $T$. The physical parameters are also present in the sample points from the SRFS-MS model. Hence, the proposed average conductivity expression directly captures the effect of these physical parameters.

## VI. VALIDATION AND ANALYSIS OF THE MODELS

### A. SRFS-MS Model Validation

To validate the SRFS-MS model, we calculate the average conductivity for a thin film of thickness $a$ (with $\kappa_b \to \infty$) and compare it with the FS-MS model from [8]. Note, no previous work has obtained the spatial resolution of conductivity with the FS-MS approach and therefore, we are limited to comparing the average conductivity. Similarly, no previous work integrates the grain-boundary effect into the FS model for rectangular interconnects. Therefore, we show the results for thin films. It may be noted that the validation of SRFS approach (without the grain boundary effect) has already been presented in [6].

Fig. 4 shows the comparisons of the average conductivity obtained from the proposed model and from [8]. For these results, we use $d_{grain} = \min(a, b)$ as in [15] We observe a close match between the two for various values of $p$ and $R$. Moreover, the effect of $p$ is prominent when $R$ is small, when $R$ increases, the effect of $p$ reduces significantly.

### B. SRFS-MS-C3 Model Validation

We now validate the SRFS-MS-C3 model by comparing the spatial profiles obtained from it with the ones obtained from the physical SRFS-MS model. This comparison is conducted for three $\kappa$ values (0.2, 1 and 5) with three $p$ values (0, 0.5 and 0.9) and three $R$ values (0, 0.5 and 0.9), thus encompassing a broad range of parameters. We compare the spatial profiles of conductivity in a square interconnect ($\kappa_a = \kappa_b = \kappa$). (Similar comparisons for rectangular wires yielded consistent conclusions, which are not included here to avoid redundancy). Fig. 5-7 shows the spatial profiles of the conductivity for

$$\begin{aligned}\sigma_{avg}(T) &= \sigma_{SRFS-MS}(0,0,T) - \big(\sigma_{SRFS-MS}(0,0,T) - \sigma_{SRFS-MS}(1,0,T)\big)\times\left[1 - \frac{\frac{\sinh(\beta_a(T))}{\beta_a(T)}}{\xi_1(\beta_a,T)}\right] - \big(\sigma_{SRFS-MS}(0,0,T) - \sigma_{SRFS-MS}(0,1,T)\big)\times\left[1 - \frac{\frac{\sinh(\beta_b(T))}{\beta_b(T)}}{\xi_1(\beta_b,T)}\right]\\ &\quad - \big(\sigma_{SRFS-MS}(0,1,T) + \sigma_{SRFS-MS}(1,0,T) - \sigma_{SRFS-MS}(0,0,T) - \sigma_{SRFS-MS}(1,1,T)\big)\times\left[1 - \frac{\frac{\sinh(\beta_a(T))}{\beta_a(T)}}{\xi_1(\beta_a,T)}\right]\times\left[1 - \frac{\frac{\sinh(\beta_b(T))}{\beta_b(T)}}{\xi_1(\beta_b,T)}\right]\end{aligned}$$

(38)

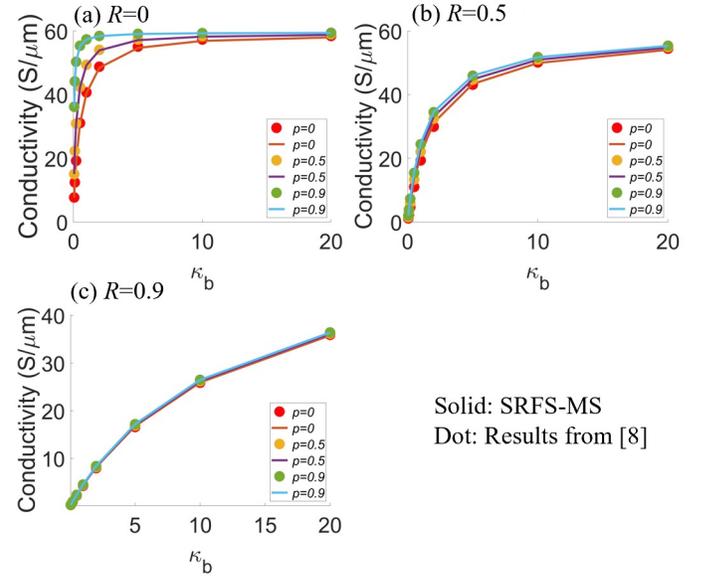

Fig. 4. Thin film average conductivity comparison for (a) $R=0$ (b) $R=0.5$ (c) $R=0.9$, showing a close match between the proposed SRFS-MS and model from [8]

TABLE I.
AVERAGE ERROR BETWEEN SRFS-MS-C3 AND SRFS-MS @ T = 25 °C

| κ | 0.2 | | | 0.5 | | | 1 | | | 2 | | | 5 | | | 10 | | |
|---|---|---|---|---|---|---|---|---|---|---|---|---|---|---|---|---|---|---|
| P / R | 0 | 0.5 | 0.9 | 0 | 0.5 | 0.9 | 0 | 0.5 | 0.9 | 0 | 0.5 | 0.9 | 0 | 0.5 | 0.9 | 0 | 0.5 | 0.9 |
| 0 | 0.1% | 0.2% | 0.1% | 0.3% | 0.3% | 0.1% | 0.5% | 0.3% | 0.1% | 0.7% | 0.3% | 0.1% | 0.3% | 0.2% | 0.1% | 0.3% | 0.2% | 0.1% |
| 0.5 | 0.1% | 0.1% | 0.0% | 0.0% | 0.1% | 0.0% | 0.1% | 0.1% | 0.0% | 0.2% | 0.1% | 0.0% | 0.1% | 0.1% | 0.0% | 0.1% | 0.1% | 0.1% |
| 0.9 | 0.0% | 0.0% | 0.0% | 0.0% | 0.0% | 0.0% | 0.0% | 0.0% | 0.0% | 0.0% | 0.0% | 0.0% | 0.0% | 0.0% | 0.0% | 0.0% | 0.0% | 0.0% |

TABLE II.
MAXIMUM ERROR BETWEEN SRFS-MS-C3 AND SRFS-MS @ T = 25 °C

| κ | 0.2 | | | 0.5 | | | 1 | | | 2 | | | 5 | | | 10 | | |
|---|---|---|---|---|---|---|---|---|---|---|---|---|---|---|---|---|---|---|
| P / R | 0 | 0.5 | 0.9 | 0 | 0.5 | 0.9 | 0 | 0.5 | 0.9 | 0 | 0.5 | 0.9 | 0 | 0.5 | 0.9 | 0 | 0.5 | 0.9 |
| 0 | 0.9% | 1.6% | 3.9% | 1.4% | 2.1% | 4.0% | 2.3% | 2.7% | 4.2% | 3.6% | 3.5% | 4.4% | 5.0% | 4.6% | 5.0% | 6.1% | 5.3% | 6.0% |
| 0.5 | 0.3% | 0.3% | 1.3% | 0.1% | 0.5% | 1.4% | 0.3% | 0.8% | 1.4% | 1.0% | 1.1% | 1.5% | 1.6% | 1.5% | 1.7% | 2.0% | 1.8% | 2.0% |
| 0.9 | 0.1% | 0.0% | 0.2% | 0.0% | 0.1% | 0.2% | 0.0% | 0.1% | 0.2% | 0.1% | 0.2% | 0.2% | 0.3% | 0.2% | 0.3% | 0.3% | 0.3% | 0.3% |

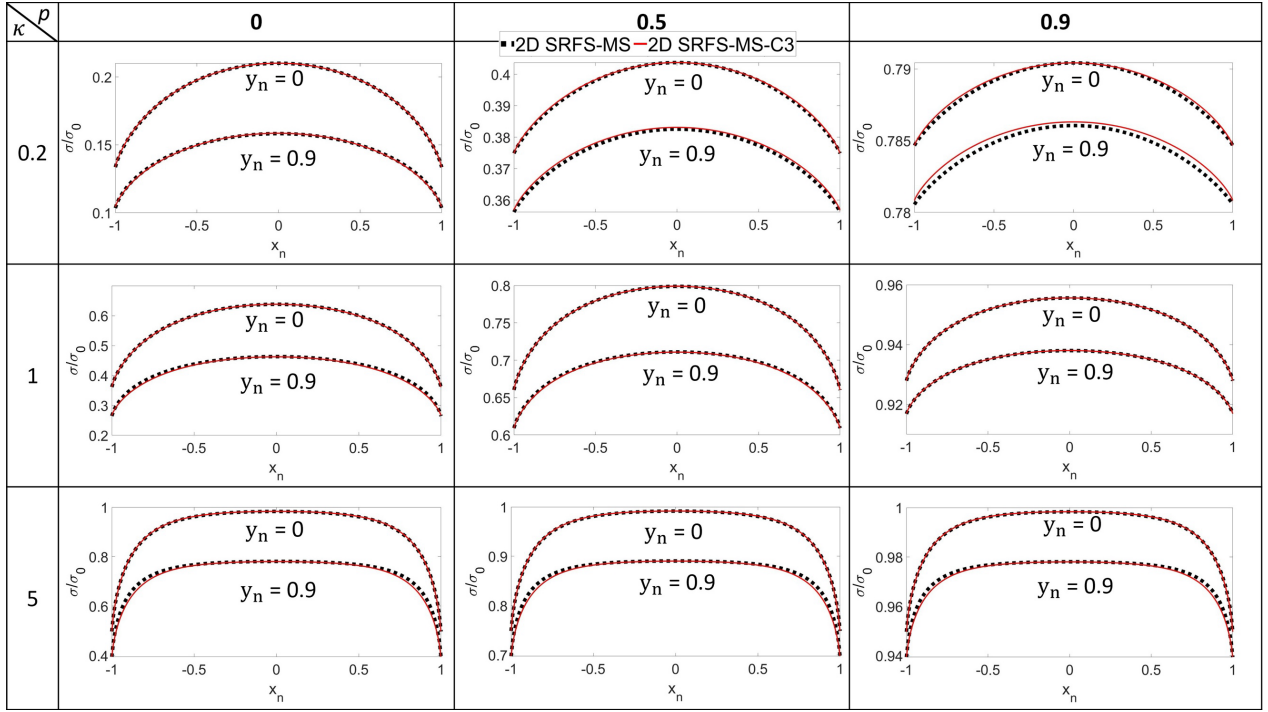

Fig. 5. The spatial profiles for $\sigma/\sigma_0$ (conductivity normalized to bulk conductivity) versus $x_n$ ($x_n=x/(a/2)$) for $y_n$ ($y_n=y/(b/2)$) = 0, 0.9 for different $\kappa$ and $p$ for $R=0$.

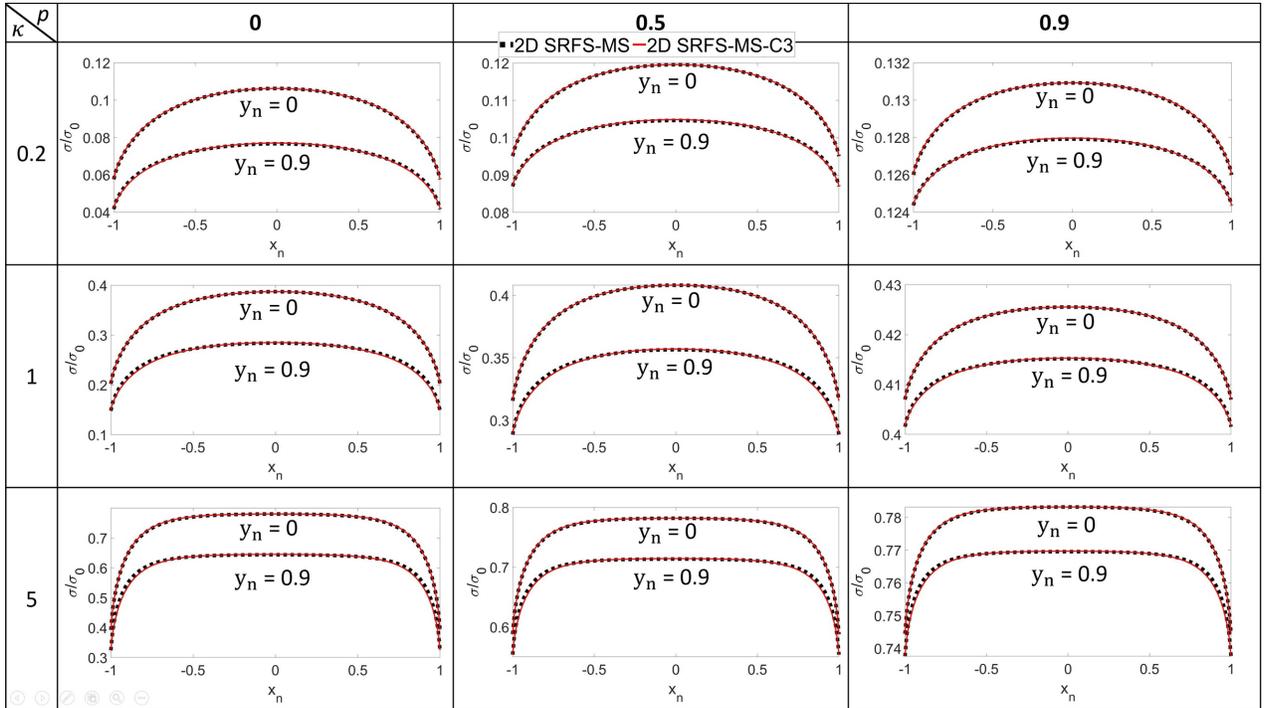

Fig. 6. The spatial profiles for $\sigma/\sigma_0$ (conductivity normalized to bulk conductivity) versus $x_n$ ($x_n=x/(a/2)$) for $y_n$ ($y_n=y/(b/2)$) = 0, 0.9 for different $\kappa$ and $p$ for $R=0.5$.

various values of $p$ and $R$. We see a close match between the two models. We also tabulate the average and maximum errors between the two models considering the spatial profile in the entire cross-section of the interconnect (Tables I and II). The average error for the SRFS-MS-C3 model is less than 0.7%, with a maximum error of less than 6.1%.

We validate the SRFS-MS-C3 with the physical SRFS-MS model for different $T$ in Fig. 8, again showing a close match. For the parameters of the $T$-dependent models of (27) and (28), we use the following from [16] based on the calibration of the model for copper interconnects with experiments: $c_0\lambda_r = 50$ S/nm, $\Theta = 343K$ and $\frac{c_L}{c_0^2\bar{v}^2} = 1.75 \times 10^{-7} \frac{J^2 n\Omega \cdot nm}{C^2 Hz}$. Tables III to VI show that the average and maximum errors for the SRFS-MS-C3 model across the two temperatures considered is less than 0.5% and 5.8% for 0°C, 0.6% and 6% for 100°C, respectively.

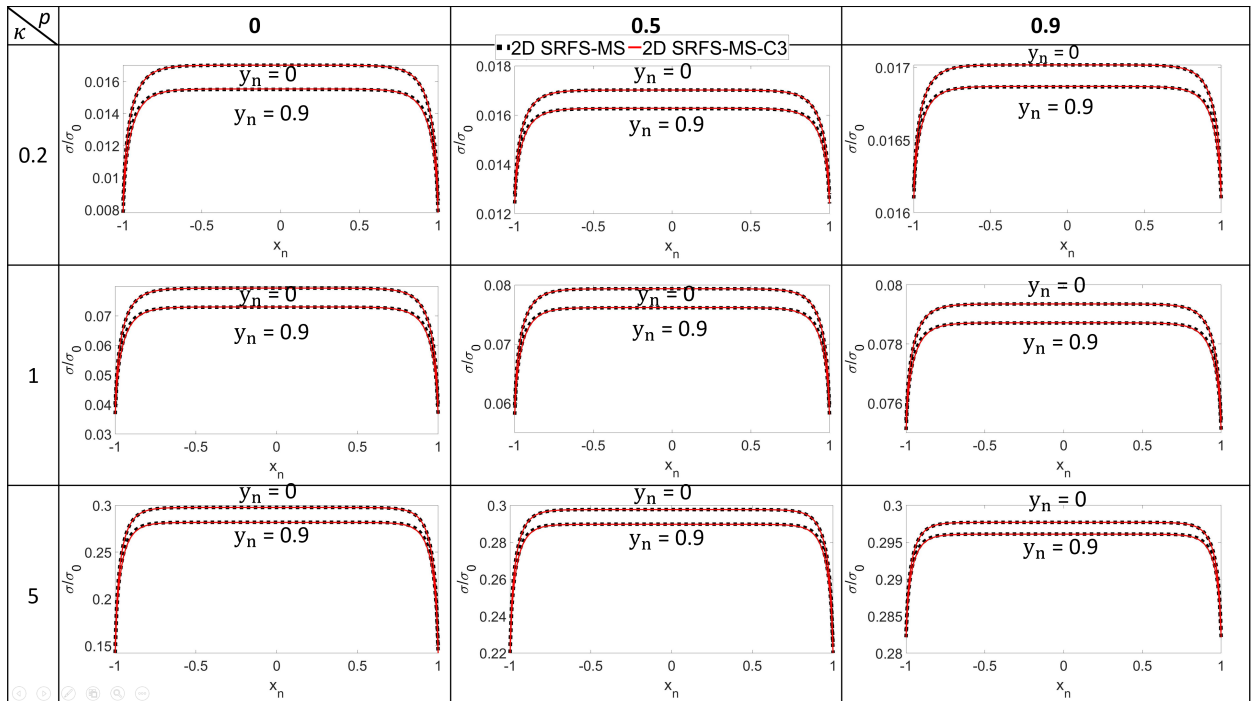

Fig. 7. The spatial profiles for $\sigma/\sigma_0$ (conductivity normalized to bulk conductivity) versus $x_n$ ($x_n=x/(a/2)$) for $y_n$ ($y_n=y/(b/2)$) = 0, 0.9 for different $\kappa$ and $p$ for $R$=0.9.

TABLE III.
AVERAGE ERROR BETWEEN SRFS-MS-C3 AND SRFS-MS @ T = 0°C

| w \ p | 7 | | | 40 | | | 200 | | |
|---|---|---|---|---|---|---|---|---|---|
| R | 0 | 0.5 | 0.9 | 0 | 0.5 | 0.9 | 0 | 0.5 | 0.9 |
| 0 | 0.1% | 0.3% | 0.2% | 0.5% | 0.4% | 0.2% | 0.3% | 0.2% | 0.2% |
| 0.5 | 0.2% | 0.1% | 0.1% | 0.1% | 0.2% | 0.1% | 0.1% | 0.1% | 0.1% |
| 0.9 | 0.0% | 0.0% | 0.0% | 0.0% | 0.0% | 0.0% | 0.0% | 0.0% | 0.0% |

TABLE IV.
AVERAGE ERROR BETWEEN SRFS-MS-C3 AND SRFS-MS @ T = 100°C

| w \ p | 7 | | | 40 | | | 200 | | |
|---|---|---|---|---|---|---|---|---|---|
| R | 0 | 0.5 | 0.9 | 0 | 0.5 | 0.9 | 0 | 0.5 | 0.9 |
| 0 | 0.2% | 0.3% | 0.2% | 0.6% | 0.4% | 0.2% | 0.3% | 0.2% | 0.2% |
| 0.5 | 0.1% | 0.1% | 0.1% | 0.2% | 0.2% | 0.1% | 0.1% | 0.1% | 0.1% |
| 0.9 | 0.0% | 0.0% | 0.0% | 0.0% | 0.0% | 0.0% | 0.0% | 0.0% | 0.0% |

TABLE V.
MAXIMUM ERROR BETWEEN SRFS-MS-C3 AND SRFS-MS @ T = 0°C

| w \ p | 7 | | | 40 | | | 200 | | |
|---|---|---|---|---|---|---|---|---|---|
| R | 0 | 0.5 | 0.9 | 0 | 0.5 | 0.9 | 0 | 0.5 | 0.9 |
| 0 | 0.9% | 2.0% | 5.1% | 2.2% | 3.0% | 5.3% | 4.9% | 4.8% | 5.8% |
| 0.5 | 0.4% | 0.5% | 1.7% | 0.2% | 0.9% | 1.8% | 1.6% | 1.6% | 2.0% |
| 0.9 | 0.1% | 0.1% | 0.3% | 0.0% | 0.1% | 0.3% | 0.3% | 0.3% | 0.3% |

TABLE VI.
MAXIMUM ERROR BETWEEN SRFS-MS-C3 AND SRFS-MS @ T = 100°C

| w \ p | 7 | | | 40 | | | 200 | | |
|---|---|---|---|---|---|---|---|---|---|
| R | 0 | 0.5 | 0.9 | 0 | 0.5 | 0.9 | 0 | 0.5 | 0.9 |
| 0 | 0.9% | 2.1% | 5.2% | 2.8% | 3.4% | 5.3% | 5.4% | 5.2% | 6.0% |
| 0.5 | 0.9% | 0.5% | 1.8% | 0.6% | 1.1% | 1.8% | 1.8% | 1.7% | 2.0% |
| 0.9 | 0.1% | 0.6% | 0.3% | 0.1% | 0.2% | 0.3% | 0.3% | 0.3% | 0.3% |

*C. Analysis of the Spatial Conductivity Profiles for different p, R, $\kappa_{a/b}$ and T*

Let us now discuss the trends in the spatial profiles of the conductivity with respect to different physical parameters. First, we observe that increasing $\kappa_a$ (or $\kappa_b$) leads to sharper changes near the edges and flat profiles near the middle. Low values of $\kappa_a$ (or $\kappa_b$) show smoother profiles (as also discussed in [6], [7]). The effect of $p$ is mainly to change the magnitude of the conductivity (lower $p$ yields lower conductivity values) but have a small effect on the shape of the spatial profile.

Since our model integrates the effect of grain-boundary scattering into the surface scattering models, we next analyze the spatial profile for various $R$ values. We observe that variations in $R$ notably influences the conductivity spatial profile. Specifically, as $R$ increases, the overall conductivity decreases and the profiles become flatter in the center and sharper at the edges, akin to high $\kappa_a$ spatial profiles.

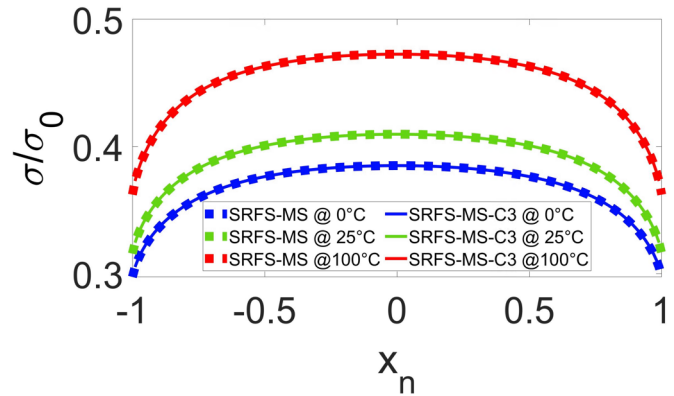

Fig. 8. The spatial profiles for $\sigma/\sigma_0$ versus $x_n$ ($x_n=x/(a/2)$) for $y_n$ ($y_n=y/(b/2)$) =0, $a=b=40$nm $p=0.5$ and $R=0.5$ for SRFS-MS and SRFS-MS-C3 for three different temperatures showing close match between them.

Mathematically, this is because $\tau^* \leq \tau \Rightarrow H(\theta) \geq 1$. As $H(\theta)$ is multiplied by $\kappa_a$ (and $\kappa_b$) terms in the arguments of the exponential and cosh functions, the effect of increasing $H(\theta)$ is similar to increasing $\kappa_a$. Physically, reduction in the relaxation time due to grain boundary scattering makes the effect of surface scattering on the spatial profile less significant as the *effective* $\lambda = \lambda_{0,EFF} = \tau^*\tilde{v}$ (instead of $\lambda_0 = \tau\tilde{v}$) becomes relatively lower in magnitude in comparison to the dimensions of the conductor.

Let us now discuss the spatial conductivity profiles for $T$ = 0°C, 25°C and 100°C for $p$ =0.5, $R$ =0.5 and $a = b$ =40 nm in Fig. 8. The main effect of increasing temperature is a decrease in the conductivity throughout the cross-section. We also observe mild changes in shape of the conductivity profile, especially for low $R$. As $T$ increases, $\lambda_0$ decreases, which makes the profile slightly sharper near the edges. (We notice this effect more prominently as the range of $T$ is increased – not shown here). As $R$ increases, the impact of $T$ on the shape of the spatial profile reduces. This is because the second term in $H(\theta)$ in (10) (which is added to 1), when multiplied by $\kappa_a$ and $\kappa_b$ in (9), yields a $T$-independent term. (This can also be understood by looking at (30) and (31)). This term reduces the impact of $T$ on the spatial profile. Thus, the incorporation of the GB scattering integrated with surface scattering has important implications in terms of reduced sensitivity of the spatial profile on temperature.

*D. Effect of integrating the grain boundary and surface scattering*

Let us further analyze how integrating the grain boundary scattering effect within the surface scattering model impacts the space-dependent conductivity and average conductivity. For that, we compare the SRFS-MS model with the standard Matthiessen's approach (which is equivalent to setting $H(\theta)$ to 1 in our model). Fig. 9 shows the comparisons. We make note of a couple of observations. First, the conductivity predicted by Matthiessen's approach is lower than if the GB scattering is integrated within the surface scattering. This is because of the $H(\theta)$ term (which is > 1 for $R > 0$, and = 1 for $R$=0) in the latter approach. Since $H(\theta)$ is the denominator of $\Delta\sigma_{SS \leftarrow GB}$, it lowers the conductivity degradation due to surface scattering for $R > 0$. Second, the spatial conductivity profile for the Matthiessen's approach is less sharp near the edges. This is because $H(\theta)$ term, multiplied by $\kappa_a$ and $\kappa_b$ in (9), effectively reduces $\lambda_{0,EFF} = (\lambda_0/H(\theta))$, which has a similar effect on the spatial profile as increasing $\kappa_a$. This makes the spatial profile sharper near the edges (compared to the Matthiessen's approach). In addition to these two effects, we have already discussed how the incorporation of $H(\theta)$ affects the temperature dependence.

## VI. CONCLUSION

In this paper, we proposed a 2D spatially resolved FS and MS (SRFS-MS) based model for conductivity to physically capture surface scattering and grain boundary scattering in interconnects with rectangular cross-sections. The SRFS-MS model extends the SRFS model proposed in [6], [7] by integrating the effect of grain boundary scattering within the surface scattering model via the $H(\theta)$ function. This model retains all the advantages of the SRFS model, including spatial dependence of conductivity and direct relationships with physical parameters. With the integration of grain boundary effects, the overall resistivity is no longer a simple addition of surface and grain boundary scattering terms as per Matthiessen's rule; instead, these two scattering mechanisms are interdependent, which is a more physically accurate representation based on [8]. We also incorporate the temperature($T$)-dependence in our SRFS-MS model. Furthermore, we introduce a circuit compatible version of the conductivity model, termed SRFS-MS-C3, for rectangular interconnects. The proposed model extends the SRFS-C3 model by employing an analytical function (explicitly related to physical parameters $p$, $R$, $\lambda_0$, $T$ and in the wire cross-sectional geometry parameters) to predict the spatial profile using only four conductivity values from the physical SRFS-MS model. Our SRFS-MS-C3 model closely aligns with the physical SRFS-MS model across a wide range of specularity, grain boundary reflectance coefficients, electron mean free paths, temperature and interconnect dimensions. Using the model, we show how $H(\theta)$ (i.e. treating the GB and surface scattering as inter-dependent) affects the spatial profile of conductivity and temperature-dependence. The proposed SRFS-MS models predict the conductivity across the cross-section of the interconnect. Thus, these models (especially the C3) can be implemented in a finite element simulation framework such as [11] to model the overall interconnect/via resistance, which is a subject of future research.

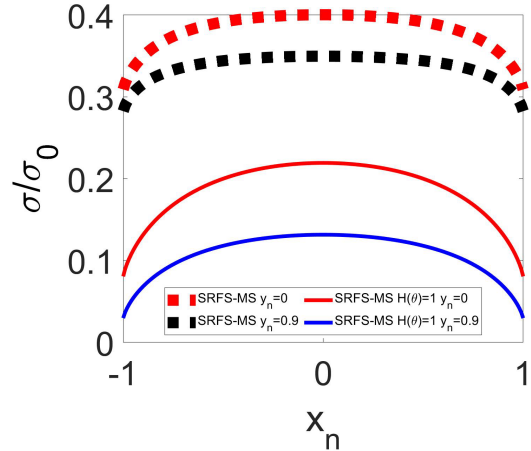

Fig. 9. The spatial profiles for $\sigma/\sigma_0$ versus $x_n$ ($x_n$=x/(a/2)) for $y_n$ ($y_n$=y/(b/2)) = 0, 0.9, $\kappa_a = \kappa_b$ =1, $p$=0.5 and $R$=0.5 showing the comparison between SRFS-MS and Matthiessen's rule (SRFS-MS with $H(\theta) = 1$). The results show that the conductivity is higher for SRFS-MS model, and that the SRFS-MS model has sharper profile near the edges.